# *Il primo solstizio d'inverno alla meridiana di Santa Maria degli Angeli in Roma*

Costantino Sigismondi*

**Sunto:** *La linea meridiana nella Basilica di Santa Maria degli Angeli in Roma fu costruita, per volontà del papa Clemente XI, tra il 1701 e il 1702 con lo scopo di misurare l'obliquità dell'asse terrestre nei successivi otto secoli. Durante il solstizio d'inverno del 1701 le prime misure utili per calcolare l'obliquità furono realizzate da Francesco Bianchini, l'astronomo che progettò la linea meridiana migliorando lo strumento simile realizzato da Giandomenico Cassini in San Petronio a Bologna. In questo articolo l'accuratezza dei dati registrati da Francesco Bianchini è discussa e comparata con le moderne effemeridi. Si presenta anche l'attuale situazione di questo storico strumento.*

**Abstract:** *The great meridian line in the Basilica of Santa Maria degli Angeli in Rome was built in 1701/1702 with the scope to measure the Obliquity of the Earth's orbit in the following eight centuries, upon the will of pope Clement XI. During the winter solstice of 1701 the first measurements of the obliquity have been realized by Francesco Bianchini, the astronomer who designed the meridian line, upgrading the similar instrument realized by Giandomenico Cassini in San Petronio, Bononia. In this paper the accuracy of the data observed by Francesco Bianchini is discussed and compared with up-to-date ephemerides. The modern situation of this historical instrument is also presented.*

---

* ITIS G. Ferraris e ICRANet, Roma; Observatório Nacional Rio de Janeiro; sigismondi@icra.it.

## 1. Arte e Scienza nella basilica di Santa Maria degli Angeli in Roma: la meridiana Clementina

La basilica di Santa Maria degli Angeli sorge nelle Terme di Diocleziano per volere del papa Pio IV e su disegno di Michelangelo (1564), il quale ha adattato la più grande aula voltata romana ancora esistente a Roma a transetto di una chiesa con pianta a croce greca.

L'orientamento di questa di questa croce è a Sud-Ovest poiché ricalca quello originale delle Terme, la cui idea originale è attribuita ad Apollodoro di Damasco che collocava il calidarium nel lato più caldo della giornata, cioè proprio il Sud-Ovest, per ridurre i tempi del riscaldamento.

Oggi proprio sulla base di queste considerazioni e sul ritrovamento delle condotte per l'aria calda sotto il pavimento del Lucernario d'ingresso, si ritiene che l'ingresso della Basilica coincidesse con il *calidarium*. Il *frigidarium* era invece una grande piscina collocata dove sta l'attuale chiostro michelangiolesco, sede del Museo Nazionale Romano delle Terme di Diocleziano.

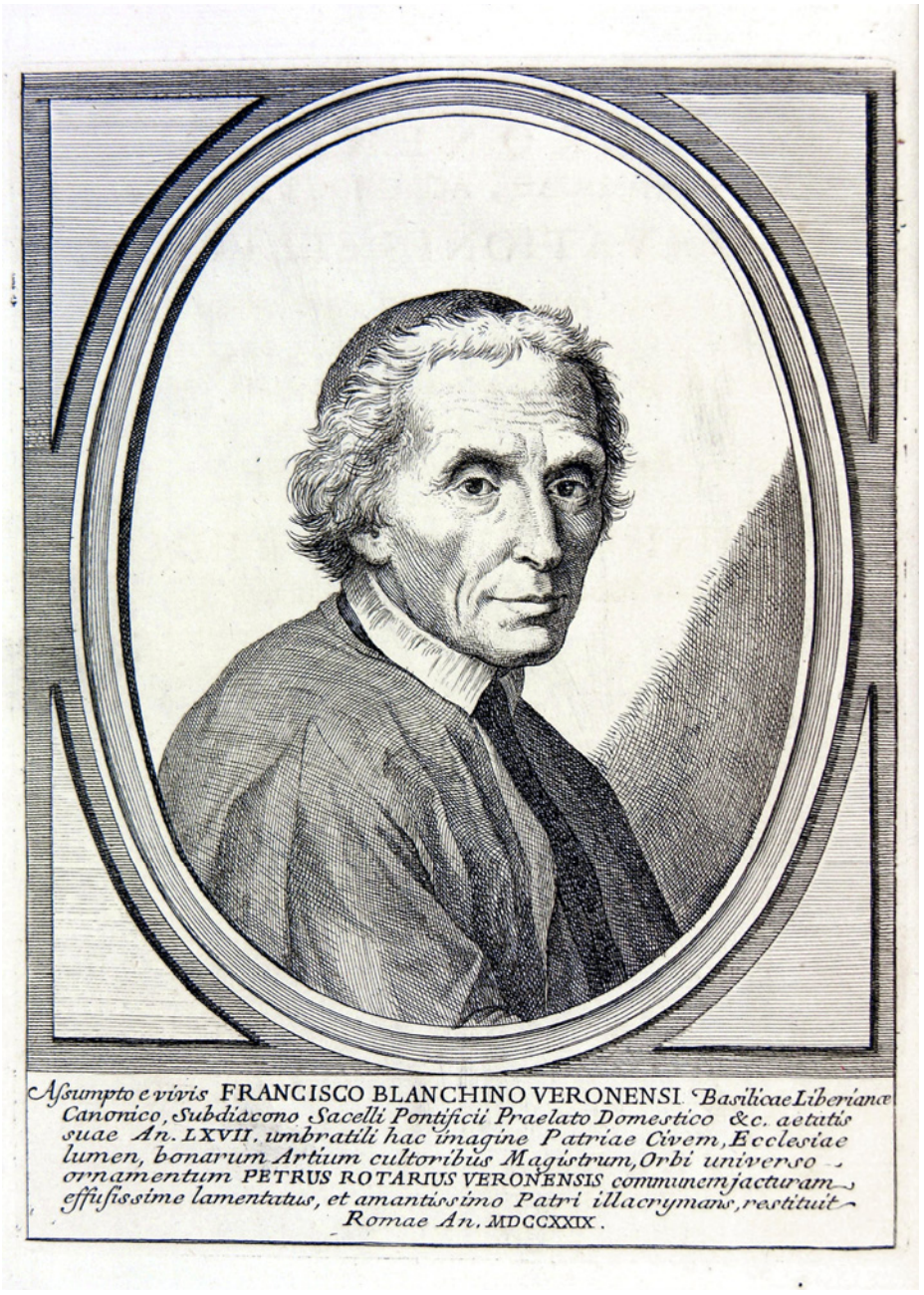

Fig. 1 - Francesco Bianchini in una stampa dell'epoca.

Fino al 1750 la Basilica appariva senza i *marmi che ornano il* pavimento e le trabeazioni di Luigi Vanvitelli, volute da Benedetto XIV per quel giubileo.

Sola, inserita tra i mattoni di cotto fiorentino spiccava, dal 1702, una striscia di marmo bianco Pentelico di circa 1 metro di larghezza e 45 metri di lunghezza, circondata da marmo giallo di Numidia, accompagna-

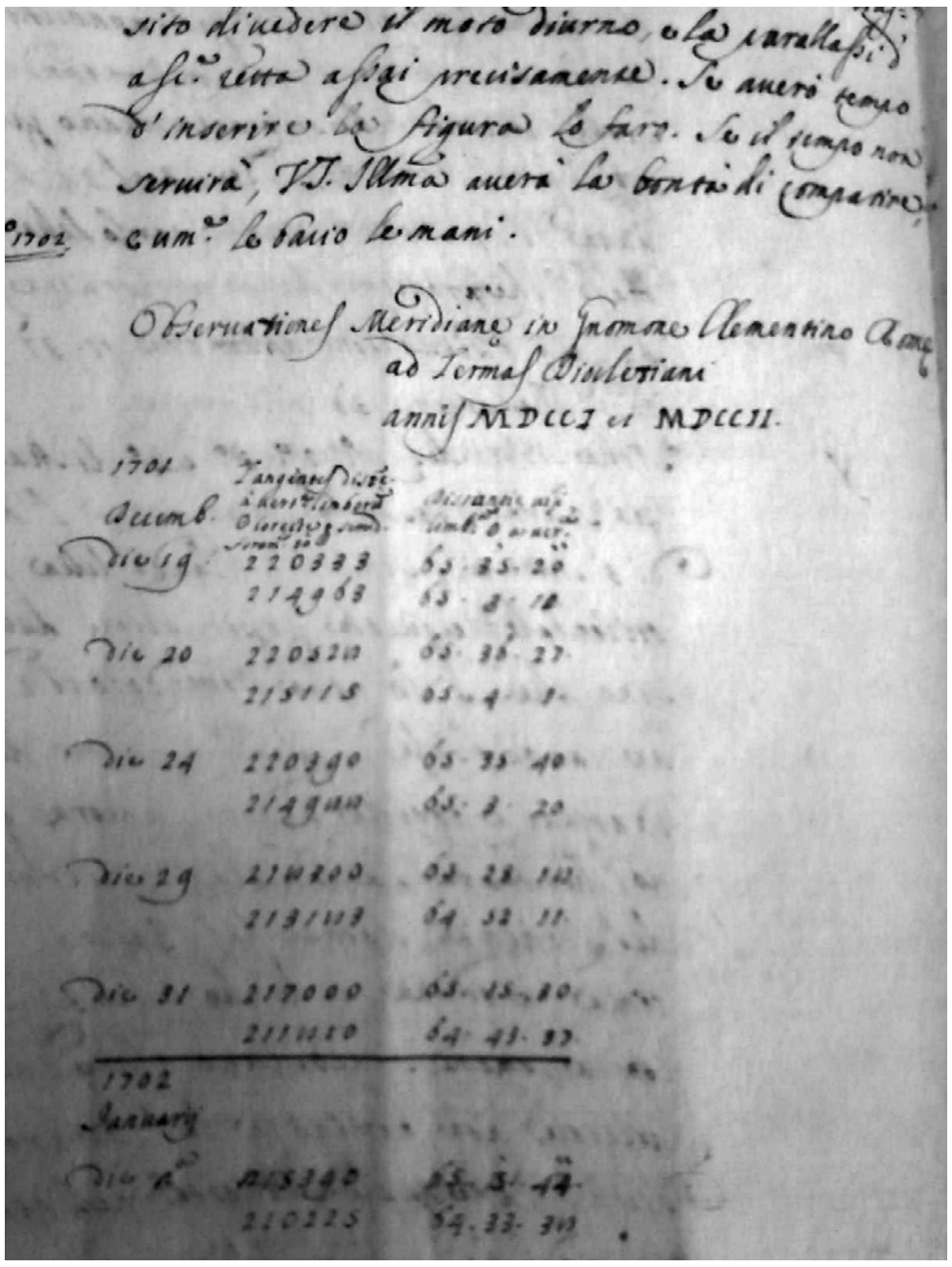

sito di vedere il moto diurno, e la parallassi
asso.te certa assai precisamente. Se averò tempo
d'inserire la figura la farò. Se il tempo non
servirà, V.S. Ill.ma averà la bontà di compatire
1702 cum.te le bacio le mani.

Observationes Meridianae in Gnomone Clementino Rome
ad Termas Diocletiani
annis MDCCI et MDCCII.

| 1701 Decemb. | Tangentes [illegible] | Distantiae [illegible] |
|---|---|---|
| die 19 | 220333 | 65. 35. 20 |
| | 214963 | 65. 8. 18 |
| die 20 | 220520 | 65. 36. 27 |
| | 215115 | 65. 4. 1 |
| die 24 | 220340 | 65. 35. 40 |
| | 214900 | 65. 2. 20 |
| die 29 | 218800 | 65. 28. 14 |
| | 213109 | 64. 52. 11 |
| die 31 | 217000 | 65. 15. 10 |
| | 211110 | 64. 41. 33 |
| 1702 Januarij | | |
| die 6 | 215340 | 65. 5. 44 |
| | 210225 | 64. 33. 30 |

*Fig. 2 - Dati solstiziali nel manoscritto di Francesco Bianchini.*

ta da 12 riquadri con le raffigurazioni dei segni zodiacali nello stile delle tarsie marmoree di pregevolissima scuola. Si trattava della meridiana Clementina, costruita per volere del papa Clemente XI, Gianfrancesco Albani (23 novembre 1700-19 marzo 1721). Della sua costruzione fu incaricato l'astronomo Francesco Bianchini (1662-1729), il quale perfezionò lo schema che Giandomenico Cassini realizzò nella meridiana di San Petronio nel 1655.

Bianchini, che era il responsabile delle antichità romane dello Stato della Chiesa, poté anche utilizzare del marmo imperiale, altrimenti introvabile a quei tempi.

Le osservazioni astronomiche per determinare la latitudine durarono dal 1° all'8 gennaio 1701, e per la fine di quell'anno la meridiana era già parzialmente funzionante, almeno per quanto riguarda la zona del solstizio invernale in Capricorno, come attestano i documenti che andremo ad analizzare nel seguito dell'articolo.

La meridiana fu visitata da papa Clemente XI il 20 agosto e il 6 ottobre 1702 in occasione delle feste di San Bernardo e San Bruno, fondatori dei Certosini.

Il legame tra papa Clemente e Santa Maria degli Angeli, la Certosa di Roma, risaliva al 6 ottobre 1700, quando, prima del conclave che seguì la morte di Innocenzo XII, appena consacrato sacerdote (27 settembre) vi celebrò la sua prima messa solenne.

Dunque la volontà di costruire una linea meridiana in Santa Maria degli Angeli era propria del nuovo pontefice; forse proprio nell'intento di rafforzare il legame tra scienza e fede, a settant'anni dal "caso Galileo".

La scelta della Basilica alle Terme per costruire uno strumento di questo genere risolveva anche uno dei problemi riscontrati già nel 1695 sulla meridiana di San Petronio a Bologna dallo stesso Cassini padre, dal figlio Jacques e da Domenico Guglielmini: l'instabilità del foro stenopeico, ovvero l'obbiettivo circolare da cui entrano i raggi del Sole e ne disegnano l'immagine sul pavimento.

La precisione necessaria per le misure della variazione dell'inclinazione (obliquità) dell'asse terrestre sul piano dell'orbita (eclittica) non permetteva nemmeno le piccole variazioni stagionali dovute alla fluttuazione della temperatura della muratura, amplificate dalla struttura delle volte ancora in assestamento.

A Santa Maria degli Angeli il foro era stato praticato in una nicchia scavata nelle mure dioclezianee, ormai prive di ogni movimento di stabilizzazione per la vetustà di ben 1500 anni.

Ragioni tecniche e affetti spirituali del papa si incontravano nella scelta di Santa Maria degli Angeli, che il veronese Francesco Bianchini, canonico di Santa Maria Maggiore, insignito degli ordini minori del lettorato e dell'accolitato, realizzò in quasi due anni di lavori.

## 2. L'arte di misurare l'immagine solare

Nella Basilica di Santa Maria degli Angeli durante le osservazioni astronomiche venivano calate delle tende esterne che coprivano i finestroni romani, come testimoniano i ganci delle stesse visibili ancora oggi all'esterno.

In questo modo la Basilica diventava un'enorme camera oscura, dove si poteva formare con il massimo contrasto possibile l'immagine del Sole, che al momento del transito meridiano si stagliava sopra il marmo bianco Pentelico, che ne favoriva la visione.

L'astronomo segnava la posizione dei due lembi superiore e inferiore dell'immagine solare, e l'assistente segnava i tempi di contatto del lembo precedente e quello seguente con il centro della linea meridiana materializzata da una lista di ottone.

Ancora oggi è possibile ripetere questa procedura, anche se la Basilica resta illuminata dai finestroni e il contrasto non arriva al livello delle osservazioni settecentesche.

I tempi di questi fenomeni possono essere rilevati con una precisione molto maggiore grazie alle tecniche video che sono state sviluppate dallo scrivente nel corso degli ultimi dodici anni, sulla base di analoghe tecniche in uso nell'astrometria solare con transiti meridiani e durante eclissi di Sole. È già stata provata l'accuratezza raggiungibile con queste riprese video, dell'ordine di un decimo di secondo di tempo, e con il metodo di segnare i bordi Nord e Sud (inferiore e superiore) del Sole che raggiunge il secondo d'arco in declinazione.

In questo articolo presentiamo la prova che questa precisione era la stessa già nel primo anno di funzionamento della meridiana, ancora prima dell'inaugurazione con il papa Clemente XI il 6 ottobre 1702.

## 3. Il primo solstizio d'inverno alla meridiana: 1701

In calce a una lettera che Bianchini scrisse al papa Clemente XI nel 1702 sono riportati i dati delle osservazioni compiute durante il solstizio invernale del 1701 dal 19 dicembre 1701 al 2 gennaio 1702.

Riportando in grafico questi dati si tracciano le parabole che meglio si adattano a descrivere le intersezioni dei bordi Sud e Nord del Sole; dai loro parametri si ricavano gli istanti del solstizio d'inverno e le posizioni dei due bordi.

Il fit quadratico dà 22.17 dicembre per il bordo Sud e 21.94 dicembre per il bordo Nord, la media è 22 dicembre alle ore 00:26

UT. Le effemeridi del Bureau de Longitudes (oggi IMCCE) danno per quel solstizio 21 dicembre ore 23:35 UT in ottimo accordo con il calcolo sui dati di Bianchini.

Lo stesso fit quadratico fornisce i valori delle intersezioni Sud e Nord dell'immagine solare al momento del solstizio:

| | |
|---|---|
| Sud: | 220.595 |
| Nord: | 215.228 |

La posizione del centro del Sole calcolato al momento del solstizio si ricava scorporando prima l'effetto della rifrazione su ciascuno dei due bordi; questo effetto nell'approssimazione al primo ordine della formula di Laplace, che a quel tempo veniva chiamata "rifrazione Cassiniana".

La formula di Laplace è:

$$z' = z - 60'' \cdot \tan(z) \quad (1)$$

dove z è la distanza zenitale in gradi come sarebbe senza l'effetto atmosfera e z' è il valore rifratto dall'atmosfera.

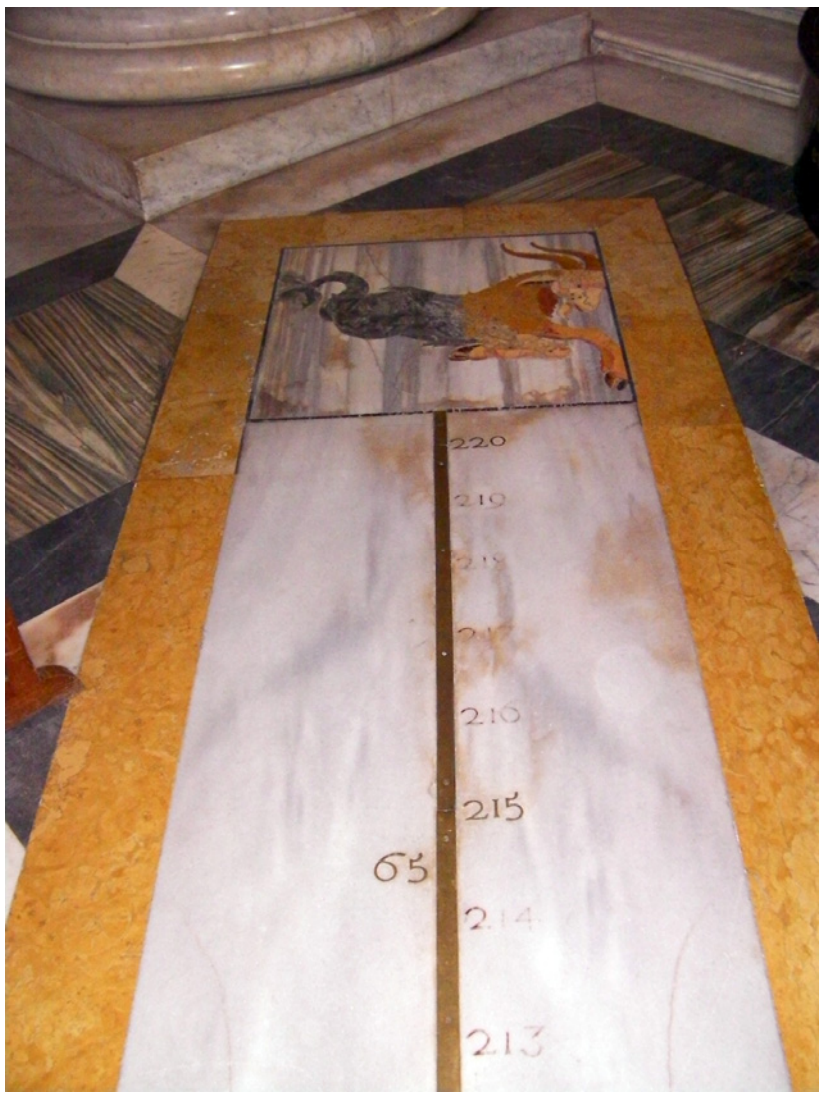

Fig. 3 - La Meridiana Clementina presso il Capricorno.

Il risultato per il nostro calcolo trova il centro del Sole a una declinazione di -23° 28' 48".

Per ottenere il corrispondente valore dell'obliquità media dell'eclittica occorre tenere ancora conto della fase della nutazione a quel tempo, che l'IMCCE calcola in $\Delta\varepsilon = -6.3''$.

Conseguentemente l'obliquità media per il 1702.0 osservata a Santa Maria degli

Angeli vale 23° 28' 54″; il valore calcolato da J. Laskar è 23° 28' 40″.

La differenza di 14″ ottenuta con la formula di Laplace che è approssimata al primo ordine, corrisponde a 7 millimetri sul pavimento della meridiana nei pressi del Capricorno.

Si noti che questa differenza è tra un valore osservato ed uno stimato da una teoria, tra le cui verifiche osservative va incluso anche questo valore osservato.

## 4. Calibrazione della meridiana

La quota del foro stenopeico determina tutta la calibrazione della meridiana, e una sua minima variazione comporterebbe uno spostamento dell'immagine solare tanto maggiore quanto più vicina fosse al solstizio invernale.

Ad esempio: un millimetro di differenza nella quota corrisponde a 2.2 mm di spostamento dell'immagine lungo la linea meridiana.

Il coefficiente di espansione lineare del calcestruzzo romano (opus coementitium) è compreso tra $\alpha \approx 10^{-5}$-$10^{-6}$/°C; considerando un'escursione termica di 30 °C tra inverno ed estate abbiamo per 20 m di altezza del foro, una variazione di quota compresa tra 1 e 6 mm.

Questa differenza può influire anche sul risultato appena trovato, in quanto la meridiana potrebbe essere stata calibrata nell'estate del 1701. Ciò implicherebbe un'altezza invernale del foro minore e la misurazione conseguente di una declinazione meno australe.

Immaginando la meridiana calibrata d'estate, la quota del foro era 20.35 m, il punto 220 della linea meridiana si trova a 44.77 m dalla verticale del centro del foro. D'inverno la quota è diventata 20.344 m (misura del febbraio 2006) e il punto 220 dovrebbe quindi trovarsi 13.2 mm più indietro.

Di conseguenza ogni misura espressa in parti centesime di altezza del foro presa d'inverno ha una sistematica differenza rispet-to al valore reale segnato sulla meridiana come se la quota del suo foro fosse esente del tutto da espansione termica.

Nel caso ipotetico massimo di 6 mm di escursione estate inverno questa differenza sarebbe di 13.2 mm sulla linea meridiana. Questo significa che leggendo 220 stiamo in realtà 13.2 mm oltre il punto 220 come sarebbe ottenuto moltiplicando 2.2 volte l'altezza del foro stenopeico invernale. Quindi sarebbe 220 +13.2/203.44=220,065, ovvero 26″ più a Sud.

Adottando per il coefficiente di espansione termica il valore estremo inferiore $\alpha \approx 1.2 \cdot 10^{-6}/°C$ si ottiene 0.72 mm di variazione di quota, e 1.6 mm di "accorciamento" della meridiana.

Le misure invernali di declinazione avrebbero così un errore sistematico di 3″: i dati corretti devono essere più australi di 3″.

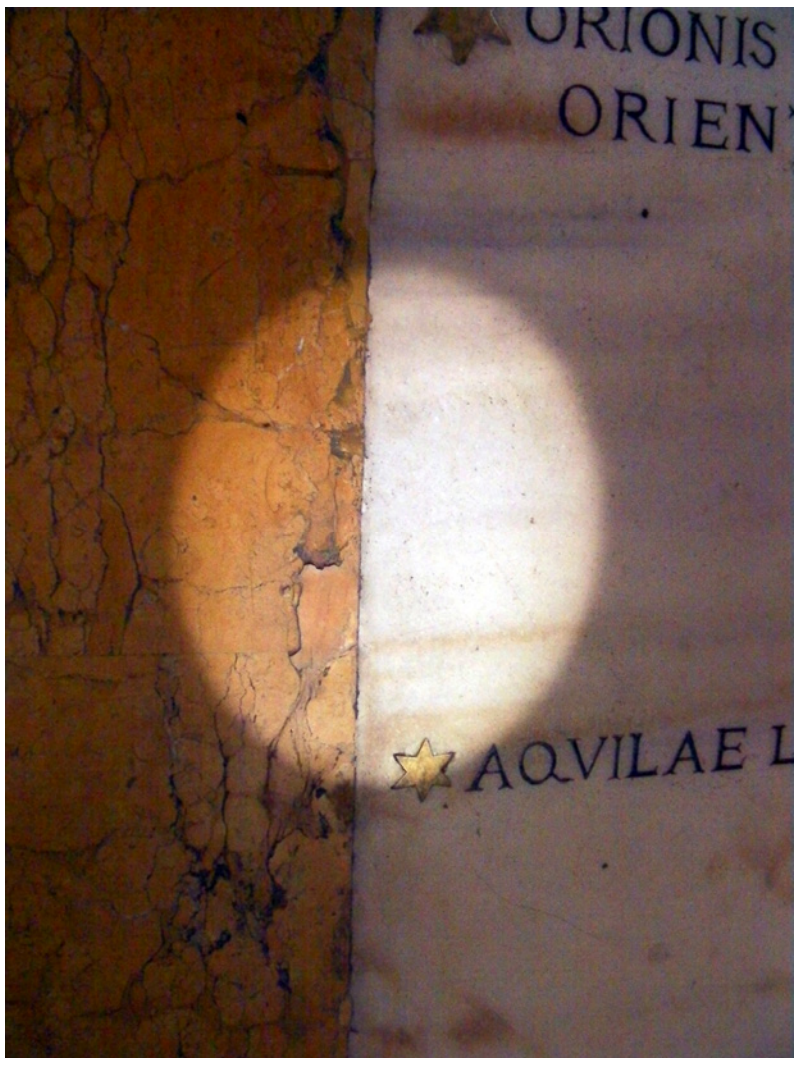

Fig. 4 - Immagine del Sole alla stessa declinazione di Altair, la ***"Lucida dell'Aquila". Foto del 2 settembre 2014.***

Questa correzione va ad aumentare di 3″ la discrepanza tra losservazione fatta da Bianchini nel dicembre 1701 e la teoria di Jacques Laskar degli anni novanta del novecento.

Manca ancora una misurazione estiva dell'altezza del foro stenopeico, con la quale sarà possibile valutare con maggiore precisione il coefficiente di espansione lineare dell'opus coementitium delle mura dioclezianee.

## Referenze

- F. Bianchini, (1702). *Lettera a Papa Clementa XI*, Roma Biblioteca Vallicelliana
- F. Bianchini, (1703). De Nummo et Gnomone Clementino, Roma
- J.Laskar, http://www.neoprogrammics.com/ obliquity_of_the_ecliptic
- C. Sigismondi, (2012). Measuring the position of the center of the Sun at the Clementine Gnomon of Santa Maria degli Angeli in Rome , arXiv 1201.0510.
- C. Sigismondi, (2006). *Pinhole Solar Monitor tests in the Basilica of Santa Maria degli Angeli in Rome, in Solar Activity and its Magnetic Origin*, Proceedings of the 233rd Symposium of the International Astronomical Union held in Cairo, Egypt, March 31 - April 4, 2006, Edited by Volker Bothmer; Ahmed Abdel Hady. Cambridge: Cambridge University Press, 2006, p.521-522
- C. Sigismondi, (2006). *Le Meridiane nella Chiesa*, Amici dei Musei 105-106, 119.
- G. Paltrinieri, (2001). *La Meridiana della Basilica di San Petronio in Bologna*, Inchiostri Associati Editore, Bologna.
- C. Sigismondi, (2010). Misura del ritardo accumulato dalla rotazione terrestre, ΔUT1, alla meridiana clementina della Basilica di Santa Maria degli Angeli in Roma, in «Mensura Caeli» Territorio, città, architetture, strumenti, a cura di M. Incerti, p. 240-248, UnifePress, Ferrara.
- www.imcce.fr
- H. Andrei, C. Sigismondi and V. Regoli, (2014). *Recent developments and prospects in ground-based and space astrometry*, in JOURNÉES 2014 SYSTÈMES DE RÉFÉRENCE SPATIO-TEMPORELS, Pulkovo Observatory, St. Petersburg, Russia 22-24 September 2014.
- V. Regoli, (2014). *Monsignor Francesco Bianchini e papa Clemente-XI: Scienza e cultura nella Roma del settecento*, Roma, tesi di Master in Scienza e Fede, Ateneo Pontificio Regina Apostolorum.